\documentclass[doublecol]{epl2} 
\usepackage{graphicx}
\usepackage{url}

\newcommand{\hx}{\hat{x}}

\title{Data reconstruction for complex flows using AI: recent progress, obstacles, and perspectives}
\shorttitle{Data reconstruction for complex flows using AI}

\author{Michele Buzzicotti \footnote{michele.buzzicotti@roma2.infn.it}}

\institute{                    
  Department of Physics and INFN, University of Rome ``Tor Vergata", Via della Ricerca Scientifica 1, 00133, Rome, Italy\\
}

\pacs{nn.mm.xx}{First pacs description}

\abstract{%In recent years significant breakthroughs in exploring big data with machine learning, (ML), have been made. 
{In recent years the fluid mechanics community has been intensely focused on pursuing solutions to its long-standing open problems by exploiting the new machine learning, (ML), approaches.} 
The exchange between ML and fluid mechanics is bringing important paybacks in both directions. The first is benefiting from new physics-inspired ML methods and a scientific playground to perform quantitative benchmarks, whilst the latter has been open to a large set of new tools inherently well suited to deal with big data, flexible in scope, and capable of revealing unknown correlations. 
{A special case is the problem of modeling missing information of partially observable systems.} {The aim of this paper is to review some of the ML algorithms that are playing an important role in the current developments in this field, to uncover potential avenues, and to discuss the open challenges for applications to fluid mechanics.}}
\begin{document}

\maketitle

\section{INTRODUCTION}

There is no doubt that our ability to produce, collect and analyze data is rapidly increasing boosted by a positive feedback loop between technological progress and new algorithms. Computer scientists, engineers, as well as physicists and mathematicians, are pushing toward the new machine learning (ML) era, which has already resulted in reforming standard data-analysis paradigms.
Breakthroughs have been achieved in numerous areas of computer science, from computer vision (CV) \cite{yeh2017semantic,ulyanov2018deep}, up to natural language processing \cite{bowman2015large, chowdhary2020natural, wolf2020transformers} as well as in some scientific contexts as the protein folding problem~\cite{jumper2021highly}. \\

In \textbf{\textit{complex flows}} as well, there have been numerous positive outcomes in nearly all testing scenarios, varying from control problems as single and multi-agents navigation in complex environments \cite{biferale2019zermelo, buzzicotti2021optimal, reddy2018glider, alageshan2020machine, verma2018efficient, calascibetta2023taming, loisy2022searching, heinonen2022optimal, loisy2023deep}, to turbulent control and drag reduction \cite{bucci2019control, park2020machine, ren2020active, buzzicotti2020statistical, huang2022machine, guastoni2023deep}, up to data assimilation problems \cite{Carrassi18, reichstein2019deep, corbetta2021deep, schultz2021can, willard2022integrating, buzzicotti2022inferring, bolton2019applications, park2019reconstruction, stock2020comparison, lou2021application, pietropolli2022gans, buongiorno2022super} to cite few of them. However, applications in fluids are still in their infancy, and the majority of cases are either conducted on highly idealized setups or only showing preliminary results on more realistic conditions. 
{The objective of this paper is to examine some of the ML tools that have been applied with promising results to reconstruct data from incomplete observations of complex systems,} including idealized turbulent~\cite{mohan2020spatio, woodward2021physics, kim2021unsupervised, fukami2021machine, buzzicotti2021reconstruction, di2022reconstructing, di2023reconstructing, yousif2023deep}, engineering~\cite{nakamura2021convolutional, guemes2021coarse, fukami2022machine, eivazi2022physics}, and geophysical flows~\cite{buongiorno2022super, pietropolli2022gans}, and to discuss the possible future directions for quantitative advancements in fluid mechanics.

\textbf{\textit{Data reconstruction}} is the art of filling in missing information by interpolating, denoising, or super-resolving a single realization, or a time series, of data fitting a specific statistical distribution~\cite{asch_data_2016,little2019statistical}. 
ML applications for data reconstruction are emerging in many areas, from computer vision~\cite{pathak2016context,zhu2017toward,belthangady2019applications,zavrtanik2021reconstruction}, to medical imaging~\cite{wang2018image,maier2019gentle,wang2020deep} up to seismic data reconstruction~\cite{chai2020deep,wangben2019deep} and astrophysics~\cite{caldeira2019deepcmb,moriwaki2020deep}. {Also in geophysical fluid dynamics works using ML to reconstruct missing data are rapidly growing~\cite{sammartino2020artificial, di2021data, brajard2021combining, shrira2020upper, fablet2021end, dong2022recent}.}
\begin{figure*}[h]
    %\hspace{-1cm}
    \centering
    \onefigure[width=1.\textwidth]{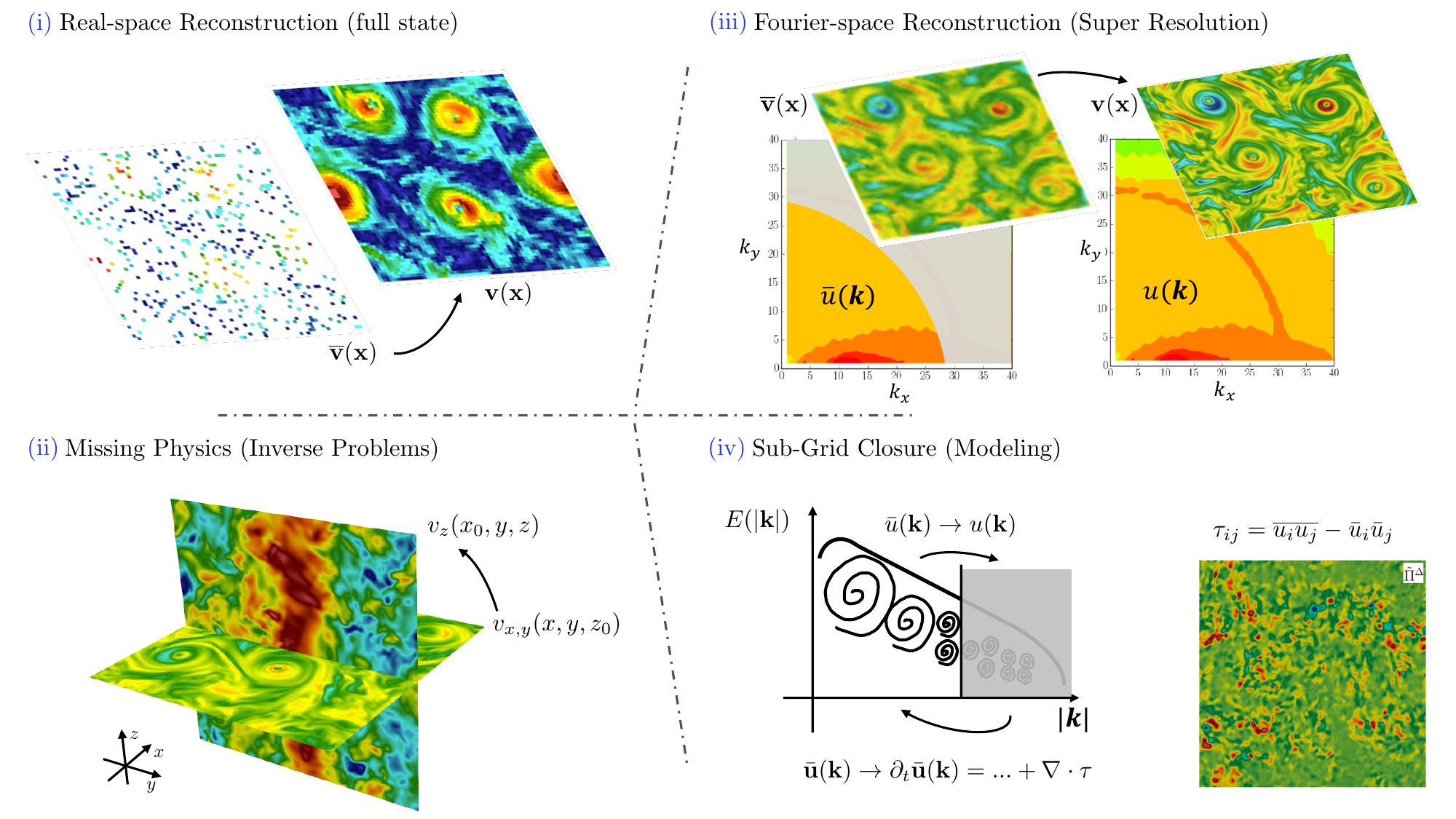}
    \caption{Graphical illustration of different types of reconstructions. Panel (i) the full state reconstruction from partial observations, gap fill. Panel (ii) inverse problems, reconstruction of missing physical quantities coupled to the observed ones. Panel (iii) Super-resolution, this is equivalent to a gap reconstruction on the high Fourier space frequencies. Panel (iv) Dynamical reconstruction, or modeling of missing physics on the observed scales.}
    \label{fig:fig1}
\end{figure*} 
For our focus on reconstructing complex flows it is possible to distinguish four possible different questions, see Fig.~\ref{fig:fig1}: \textbf{(i)} full-state restoration, with the aim to fill missing gaps in the real space state of a complex flow, \textbf{(ii)} inferring missing fields, which can be derived as the inverse problem solution where a physical observable that cannot be accessed/measured directly can be inferred by measuring other quantities to which it is coupled, \textbf{(iii)} super-resolution, which can be seen as the equivalent of point (i) but when the gap to fill is on the high-wavenumbers of the Fourier domain, \textbf{(iv)} dynamical modeling, which consists of reconstructing dynamically the effects of missing scales on the evolution of the resolved ones~\cite{meneveau2000scale, buzzicotti2018effect, biferale2019self, buzzicotti2021inertial}. 
The issue of designing a ML-inspired sub-grid closure for modeling computational fluid dynamics is a subject per se and has been recently reviewed in~\cite{duraisamy2019turbulence, vinuesa2022enhancing}. 
Here, our focus lies on the first three categories of problems under the assumption that the amount of missing information to fill in is very large, which renders the problem ill-posed. This means that multiple solutions can fit within the same reconstruction \cite{buzzicotti2021reconstruction, guemes2021coarse}. Under this assumption, already defining what the optimal solution is, it is a question that can have different answers depending on the specific target. For instance, as discussed in~\cite{li2022data, li2023generative}, the optimal reconstruction providing the minimum mean squared error (MSE) is different from optimal solutions in terms of other statistical quantities. \textit{In this review, we target reconstructions that maximize the correlations with the observed data while respecting the statistical features of the ground truth solution.}
The large-gap assumption is required when dealing with the reconstruction of complex flows. For instance, let us consider the full-state reconstruction problem of ocean surface currents. Even though satellites have allowed us to get, for the first time, a global picture of the ocean~\cite{storer2022global,buzzicotti2021coarse}, from mesoscale eddies up to western boundary currents, over time scales relevant to climatological studies (decades), observing the full dynamics of the ocean remains a gigantic task \cite{pujol2016duacs,ballarotta2019resolutions}, and requires filling gaps of spatial scales between hundred km up to less than a meter and time-frequency gaps spanning weeks up to turbulent and wave scales (of seconds), which cannot be neglected to explain turbulent stirring, mixing, and all vertical motions.

On top of applications, there are \textbf{\textit{fundamental questions}} associated with reconstructing complex flows. What type and quantity of information are required to perform different reconstructions is one open theoretical question, which can be investigated via a reverse engineering approach: different inputs can be passed to the same model to assess the impact on the reconstruction quality.
Here, we discuss {some of the ML} algorithms that are transforming the conventional paradigms of data analysis and that have the potential to facilitate breakthroughs in the field of fluid dynamics.
Following a chronological order, we start with an introduction of `Variational Auto-Encoders', (VAEs), `Generative Adversarial Networks', (GANs), and `denoising Diffusion probabilistic Models', (DMs). After we discuss how to combine pure Data-Driven methods with the physical knowledge at hand and we provide possible future directions in this discipline.

\section{DATA-DRIVEN METHODS}
\label{sect:DD-methods}

A typical approach to repair missing data in gappy fields, before the rise of ML, was based on \textbf{\textit{proper orthogonal decomposition}} (POD). POD is used to reduce data dimensionality by identifying the dominant patterns in a dataset and representing them using a smaller set of orthogonal basis functions, (POD modes, i.e. eigenvectors of the correlation matrix)~\cite{sirovich1987low, fukunaga2013introduction, hayase2015numerical}. {The same approach have been extensively used also for filling of missing points in geophysical data sets where it takes the name of Empirical Orthogonal Function~\cite{kondrashov2006spatio}.} Extension of such techniques as the Gappy POD (GPOD)~\cite{everson1995karhunen} or the Extended POD (EPOD)~\cite{maurel2001extended} were derived to repair missing data with minimal MSE solutions, showing results outperforming Kriging interpolation~\cite{gunes2006gappy}. However, POD-based approaches are limited when dealing with complex multi-scale and non-Gaussian statistics as is the case of turbulent flows. As shown in ~\cite{guastoni2021convolutional}, where they implemented EPOD to reconstruct the bulk velocity of wall-bounded turbulence from wall measurements, and in~\cite{li2022data} where they used EPOD and GPOD to fill missing data on velocity planes extracted from 3d rotating turbulent flows, {POD methods can only reconstruct the large-scale, Gaussian features of the ground truth data.}\\

\begin{figure}[h]
    \centering
    \onefigure[width=0.45\textwidth]{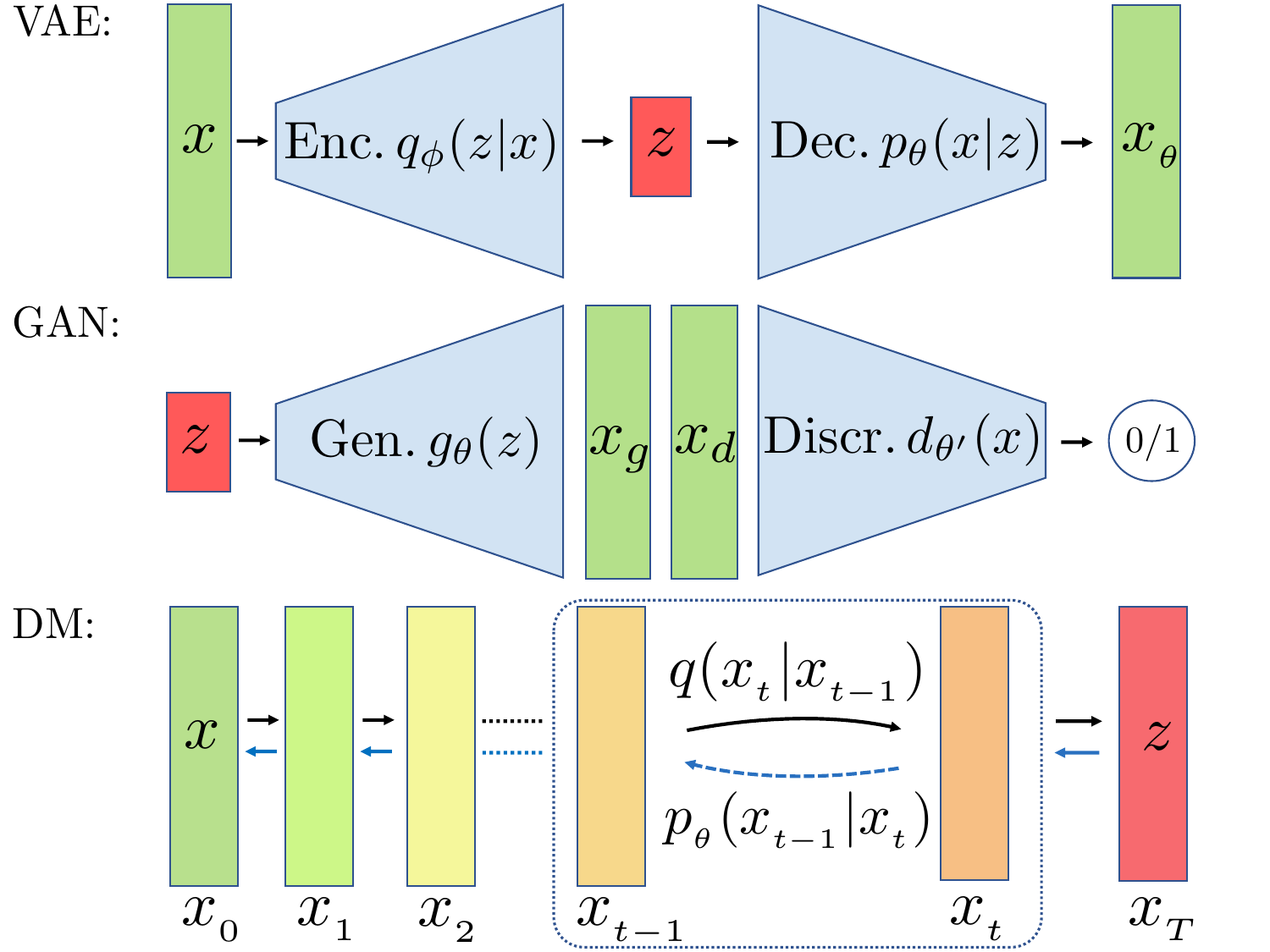}
    \caption{Schematic representation of the three ML algorithms developed to generate data accordingly with the probability distribution function described by a training dataset. (Top) Variational Auto-Encoders, based on a probabilistic encoder $q_\phi(z|x)$ and decoder $p_\theta(x|z)$. (Middle) Generative Adversarial Networks, based on a generator $g_\theta(z)$ and a discriminator $d_\theta(x)$. (Bottom) Diffusion Models, aiming to model the reverse transition probability of a Markov Chain, through a network $p_\theta(x_{t-1}|x_t)$.}
    \label{fig:fig2}
\end{figure}

A significant advancement in this regard was brought by the ML \textbf{\textit{generative models}}~\cite{hong2019generative, bond2021deep, gui2021review} aiming to generate new data that resembles `statistically' the training dataset. Their success can be attributed to two factors. Firstly, their architecture relies on multi-layer Convolutional Neural Networks (CNN)~\cite{o2015introduction, li2021survey}, which inherently possess the ability to emphasize long-range correlations in data. Secondly, they are trained with loss functions that not only account for MSE accuracy but also for statistical differences between the generated and ground truth data.
Fig.~\ref{fig:fig2} gives a schematic illustration of the three main generative models that are commonly utilized in ML. 
\textbf{\textit{VAEs}} have been the first type of neural network trained to give in output new samples, not belonging to the training dataset, which satisfies the same statistical properties. 
In the first row of Fig.~\ref{fig:fig2}, a simple diagram depicts the workflow of the VAE model. VAEs, like their predecessors, Auto-Encoders, are based on an encoder-decoder structure. However, unlike Auto-Encoders, the aim of VAEs is not to perform a dimensionality reduction by projecting the input data, $x$, into a low-dimensional latent space, $z$. Instead, VAEs define a probabilistic decoder, $p_\theta(x|z)$, which maps any input from the latent space, sampled from a simple distribution, $p(z)$, typically a multivariate Gaussian, into a sample in the output space that satisfies the (generally unknown) statistical distribution characterizing the training dataset. The probabilistic encoder, $q_\phi(z|x)$, plays a crucial role in VAEs by facilitating the sampling of the latent space, $z$, during training to accelerate decoder convergence. The encoder's primary objective is to model the posterior probability of the decoder, denoted as $p_\theta(z|x)$, which corresponds to the likelihood of obtaining a particular sample in the latent space $z$ when generating a specific input from the dataset $x$.
The presence of the probabilistic encoder assists the discriminator in exploring a smaller and more relevant sub-manifold of the latent space, resulting in faster and more stable training. VAE models operate on the basic assumption of learning a mapping between a simple and fixed distribution into the data probability distribution. Training the encoder entails minimizing the Kullback-Leibler Divergence (KLD) between the selected latent space distribution, $p(z)$, and the encoder's output distribution, $q_\phi(z|x)$. This operation can generally be computed analytically and requires adding a few extra terms to the decoder loss function.
As previously mentioned, the probabilistic decoder of the VAE is trained by maximizing the log-likelihood of the generated data, $\log p_\theta(x)$, where;  $$p_\theta(x) = \int p_\theta(x|z) p(z) dz .$$ Directly computing this loss function is intractable. However, a lower bound can be defined, using a variational inference formulation, and calculated under some approximations. The approximations are based on the assumption that the decoder's errors are Gaussian. By making this assumption, the maximization of the log-likelihood can be rewritten as a minimization of the MSE. While this approximation may be reasonable in some contexts, it is certainly unsuitable for considering turbulent flows, which are well-known for their highly non-Gaussian extreme fluctuations. As shown in the context of turbulent flows on a rotating frame~\cite{li2022data}, the minimization of MSE alone results in generating solutions that match the training data only at the large, more energetic scales, while over-damping the smaller scales. Therefore, rather than as generative methods, VAEs are mostly considered in the context of reduced-order modeling to perform a probabilistic projection on low-dimensional latent space, z, as studied in the context of 3d homogeneous and isotropic turbulent (HIT) flows~\cite{mohan2020spatio} and more recently on a 2d flow of a simplified urban environment~\cite{eivazi2022towards}.\\
\textbf{\textit{GANs}} are proposed to improve VAEs by relaxing the Gaussian errors assumption, and by improving the evaluation of statistical features in generated data in the loss~\cite{goodfellow2014generative, wang2017generative, heusel2017gans}. In general, the functional form of the probability distribution that characterizes the training dataset is unknown. To overcome this issue, a second network, the discriminator, $d_{\theta'}(x)$, is used to evaluate the statistical properties of training and generated datasets. The discriminator provides a loss function that the GAN generative part, $g_\theta(z)$, can optimize during training.
The discriminator functions as a classifier and is trained to assign a probability of an input being generated or extracted from a true dataset. On the other hand, the generator maps a sample from a latent space into a sample in the data space, similar to a VAE decoder. Its objective is to generate increasingly realistic samples that can fool the discriminator, from which comes the name `adversarial', {where, as in a zero-sum game, a gain for one network gives an equivalent loss to the other~\cite{hofbauer1998evolutionary, borra2022reinforcement}}. {For a fixed generator the analytical expression for the optimal discriminator can be derived by maximizing the adversarial loss~\cite{goodfellow2014generative}, and results in;} $$d^*(x) = \frac{p_{true}(x)}{p_{true}(x)+p_{gen}(x)} , $$ where $p_{true}$ and $p_{gen}$ represent the statistical distributions of the true and generated datasets. Similarly, the optimal generator, denoted as $g^*(z)$, can be derived as the network that minimizes the Jensen-Shannon Divergence (JSD), a symmetric formulation of the KLD, between the true and generated distributions~\cite{goodfellow2014generative}.
GANs, have exhibited unparalleled potential in producing turbulent datasets that display a remarkable level of statistical similarity to their original counterparts. Both the original and generated data exhibit identical deviations from Gaussianity up to the evaluation of high-order statistical observables in several setups, as in super-resolving to a $64\times$ larger 2d turbulent flows behind cylinders~\cite{deng2019super}, and of 3d HIT flows ~\cite{subramaniam2020turbulence, kim2021unsupervised}, as well as in filling large gaps in rotating turbulence~\cite{buzzicotti2021reconstruction} and 3d channel flows~\cite{guemes2021coarse}. 
\begin{figure}[h]
    \centering
    \onefigure[width=0.49\textwidth]{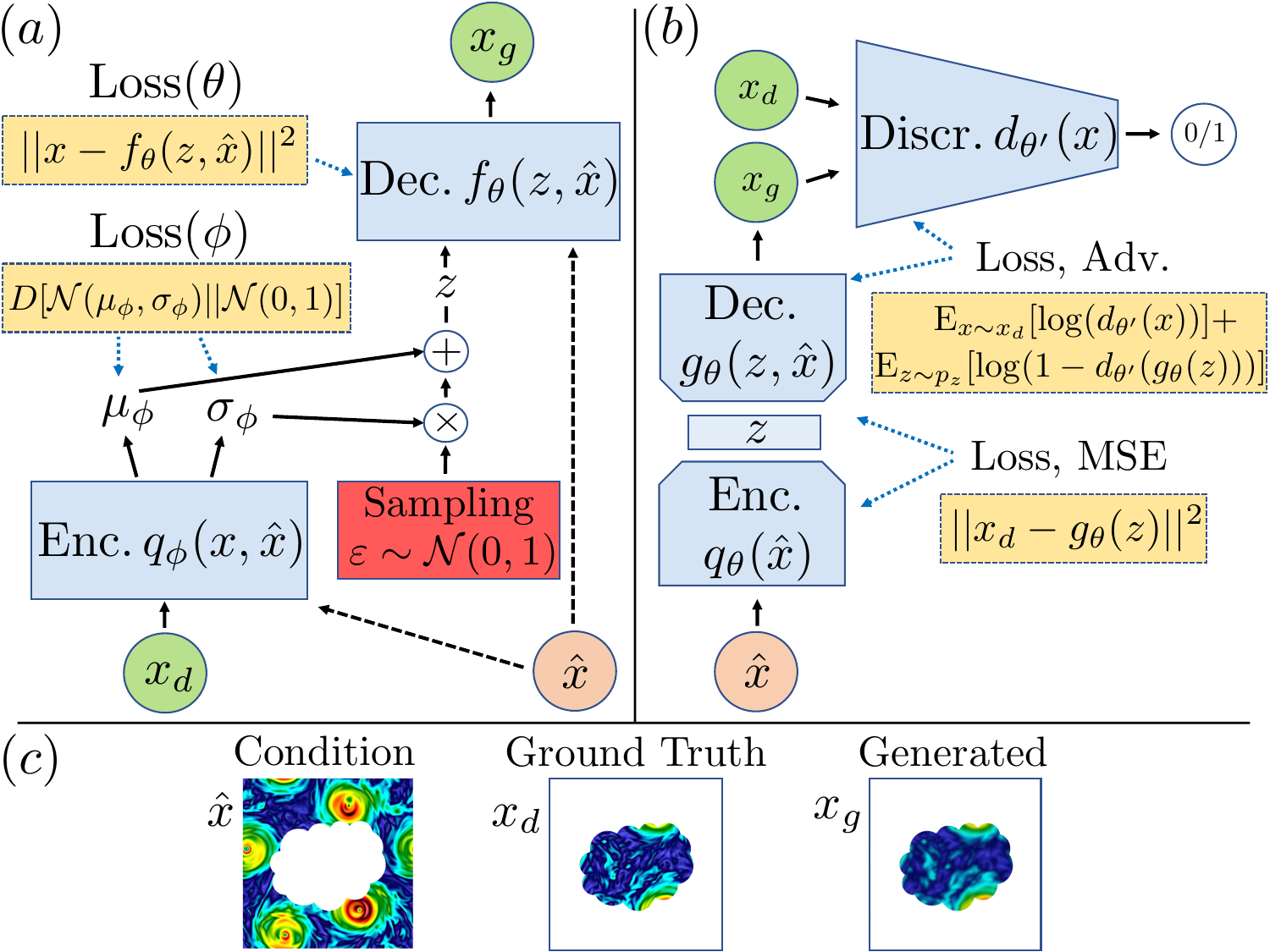}
    \caption{Workflow of a typical Variational Auto-Encoder (panel a) and of a typical Generative Adversarial Network (panel b) designed to generate samples $x_g$ conditioned on some observations $\hx$. The blue boxes represent the functions optimized during the training, the yellow boxes report the loss functions and their connection with the different parts of the network. The red boxes indicate where the stochastic sampling is happening along the network propagation, while the green and brown circles represent respectively the input and the conditioning of the networks. (panel c) Visualization of typical fields analyzed by the networks with the aim of transforming incomplete data into corresponding complete data.}
    \label{fig:fig3}
\end{figure}
Fig.~\ref{fig:fig3} showcases the workflows of VAEs and GANs focusing on the applications of these models to fill gaps by exploiting their generative capacities also when constrained to fit some observations. In the three panels (a)-(b)-(c), the sample $\hx$ represents the gappy data and serves as a condition to the model, the ground truth data (known only in the training stage) is denoted as $x_d$, the model reconstruction is called $x_g$. 
In the VAEs the condition $\hx$ is passed to both the encoder and the decoder. During training the encoder projects $x_d$ and $\hx$ into the latent space by defining the variance and the mean of a Gaussian distribution from which a sample $z$ is extracted. The loss function is the same as in the case of pure generation. In testing setup, the reconstruction the sampling on $z$ is done from a standardized multivariate Gaussian while the decoder on top of the $z$ sample analyzes also the condition $\hx$~\cite{kingma2013auto, salimans2015markov, doersch2016tutorial}. 
Panel (b) displays the GAN reconstruction setup, which distinguishes itself from the unconstrained model in that the generator employs an encoder-decoder architecture to map $\hx$ to an intermediate space $z$ prior to generating the filling data instead of performing a random sampling on the latent space. The discriminator operates as usual, but now the overall generation loss is a linear combination of the MSE between the ground truth and the reconstruction data, in addition to the adversarial loss provided by the discriminator prediction~\cite{pathak2016context, li2022data, li2023generative}.
{GAN generates realistic samples also when constrained to match some observations. However, in the reconstruction case having statistically consistent data leads to a larger MSE with respect to the ground truth solutions. Indeed a tiny shift in space between the reconstruction and the true solution brings larger MSE if the fields are both highly fluctuating~\cite{li2022data}.}
The principal limitation of GANs arises from their adversarial nature, resulting in highly unstable training and slow convergence. {It can happen that one of the two players is dominated by the other and converges into a failure solution.}\\

{\textbf{\textit{Diffusion Models}}, Fig.~\ref{fig:fig2} bottom panel, are an alternative technique to generate}. DMs transform a simple distribution into a more complex distribution, resembling the training data while avoiding the need to introduce a surrogate loss function, as seen in VAEs, and without relying on adversarial training, as in GANs.
\begin{figure}[h]
    \centering
    \onefigure[width=.49\textwidth]{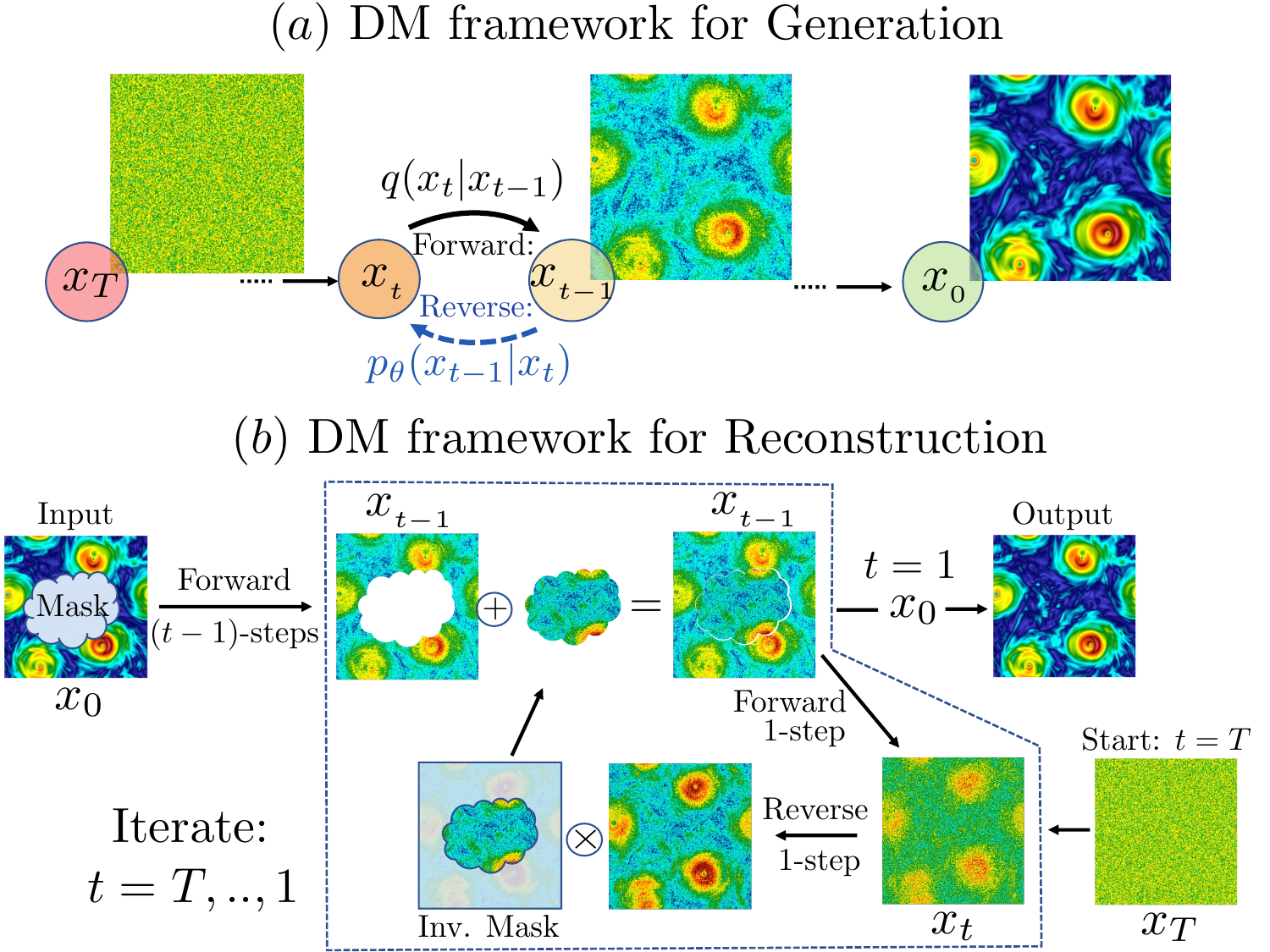}
    \caption{(Panel a) Typical workflow of a Diffusion Model designed and trained to generate data respecting the statistical distribution of a training dataset. Framework applied to reconstruct the full state (reconstructed output) from partial observations (masked input) using a pre-trained generative, unconditioned Diffusion Model starting from a random noise ($x_T$) (panel b).}
    \label{fig:fig4}
\end{figure}
The workflow of DMs is illustrated in Fig.~\ref{fig:fig4}, for both the generation (a) and the reconstruction (b) setups. 
DMs use a Markov chain to gradually convert one distribution (latent space) into another (dataset), following the idea developed in non-equilibrium statistical physics~\cite{sohl2015deep}. To learn the parameterized Markov chain, DMs are trained using variational inference to produce data samples that match the original data after repeating a finite number of steps. The learning involves estimating small perturbations to a diffusion process, problem, which is more tractable than explicitly describing the full distribution within a single jump as potentially done in the other generative models. Furthermore, since a diffusion process, $q(x_t|x_{t-1})$, exists for any smooth target distribution, this method can capture data distributions of arbitrary form~\cite{sohl2015deep}. If the forward diffusion process is a Markov chain that gradually introduces Gaussian noise to the data until the signal is destroyed, the model subsequently learns how to reverse the diffusion process and generate desired data samples starting from pure Gaussian noise realizations. Unlike VAEs or GANs, diffusion models involve a latent variable $z$ with dimensionality identical to that of the original data $x$. 
In Fig.~\ref{fig:fig4} panel (b) discusses the approach proposed in~\cite{lugmayr2022repaint} to employ pre-trained unconditional DMs to condition the generation process on filling some partial observations. The approach involves moving forward the gappy data through the Markov chain by iteratively adding noise to the observations, while simultaneously progressing backward from the noise distribution using the reverse chain learned by the DM. To incorporate the observed data in the generation process, the strategy is to repeatedly merge the forward-propagated noisy observations with the reverse-propagated noisy signal. This allows the reverse process to propagate data information within the gap and generate a correlated reconstruction. DMs have produced state-of-the-art results in image generation, see the famous example of `DALL-E 2'~\cite{saharia2022photorealistic}, demonstrating their ease of definition and effectiveness in training~\cite{ho2020denoising, nichol2021improved, dhariwal2021diffusion, lugmayr2022repaint, rombach2022high}. 
{Attention~\cite{vaswani2017attention} is another feature often implemented inside DMs architecture, which is potentially crucial for large gap-filling. Indeed, attention is meant to enhance the role of some parts of the input data while diminishing others, showing good results at handling long-range spatial relations~\cite{wang2018non}}. However, DMs and attention, have not yet been extensively applied in the generation and reconstruction of complex gappy flow data, but they have only been used to super-resolve smooth bi-dimensional Kolmogorov flows~\cite{shu2023physics}. Therefore, the investigation of DMs and `attention' capacity to generate high-quality samples of complex flows is an ongoing field of research.

\section{PHYSICS-INFORMED METHODS}
\label{sect:PIDD-methods}

Leveraging the observed data and the equation of motion, Physics-Informed techniques, exploit spatio-temporal correlations to derive accurate reconstruction of incomplete data. Kalman filters, variational approaches, and nudging are examples of advanced tools that have proven effective in enhancing initial conditions for weather forecasting problems since before ML~\cite{kalnay2003atmospheric, carrassi2018data, lakshmivarahan2013nudging}. \\
\textbf{\textit{Nudging}} is a physics-informed way to control the evolution of a flow via the continuous insertion of observed data and the addition of a penalty term, which tries to keep the flow trajectory close to that of the empirical subset~\cite{hoke1976initialization}.  Nudging has been recently applied to reconstruct high resolution HIT flow from sparse measurements~\cite{di2020synchronization} and to estimate physical unknown parameters from turbulent data~\cite{buzzicotti2020synchronizing}. While physics-agnostic ML approaches are focused solely on finding patterns in data, there is growing interest in incorporating physical knowledge into ML algorithms, particularly in the field of fluid mechanics where the underlying physical laws are well understood~\cite{karniadakis2021physics}. The first objective is to impose constraints on the ML solutions to ensure that they adhere to the known physical properties, the second objective is to streamline the training by integrating relevant information directly into the network architecture or training setup. There exist three methods for incorporating physics into ML algorithms; \textbf{(i)} observational, \textbf{(ii)} inductive, or \textbf{(iii)} learning biases. Observational biases may be introduced by selecting training data to ensure that a specific aspect of physics is not only present but also emphasized, {ie, extreme events can be shown during training more often than the frequency at which they occur.} Inductive biases embed physical constraints into the network architecture, as for example the CNNs embed invariance along the groups of symmetries possessed by typical patterns observed in images. Finally, learning biases operate in a `soft' way by adding additional terms to the loss function that penalize non-physical solutions~\cite{raissi2019physics}, such as those that do not satisfy equations of motion, violate mass or energy conservation, and so forth. Physics-informed data-driven tools have just begun to be highlighted as particularly promising in areas as numerical weather prediction~\cite{zhao2019physics, alber2019integrating, kashinath2021physics, willard2022integrating}. Improving data-driven and physics-informed methods synergy will undoubtedly be the focus of research in the upcoming years.

\section{PERSPECTIVES}
\label{sect:perspect}

Although ML techniques have been already implemented as standard tools in computer science, \textit{fluid dynamics presents challenges that differ from those tackled in many applications of machine learning, such as image recognition and advertising}, as stated in~\cite{brunton2020machine}. Fluid flows necessitate precise and quantitative evaluations of the multi-scale and multi-frequency physical mechanisms that they must adhere to. On top of this, while idealized flow setups offer large datasets of high complexity, and quality, in real-life flows, one needs to deal with very sparse and noisy data. The misalignment between the idealized cases studied in the literature, and real applications opens non-trivial problems connected with the generalizability and the uncertainty quantification (UQ) of the pre-trained models~\cite{bucci2023curriculum, abdar2021review, hullermeier2021aleatoric, barth2020dincae}.
To overcome those issues, it is highly desirable to have in the future, more open-access databases, such as JHTDB (\url{https://turbulence.pha.jhu.edu}) and
Smart-Turb (\url{https://smart-turb.roma2.infn.it}), and well-defined open challenges, { such as (\url{https://github.com/ocean-data-challenges}), that can bring different communities closer, and that can drive ML applications to go beyond theoretical exercises towards the quantitative improvement required to provide advancements in fluid mechanics.}
Today's challenges are connected with the need of a \textbf{\textit{quantitative AI}}, driven by several critical factors such as validation, benchmarks on generalization, and UQ of ML solutions. Another crucial aspect is the \textbf{\textit{problem dimensionalization}}, which involves understanding the correlation between the network's architecture, deepness, structure, and size, and the physical parameters, as Reynolds, Rayleigh, and time-to-solution, among others. 
As discussed, already defining an evaluation metric to quantify the solution quality is an issue in fluid mechanics that needs to be carefully designed.
Answering these questions is necessary, and \textbf{\textit{interdisciplinary collaborations}} between applied scientists and AI specialists, are unavoidable for establishing best practices outperforming today's data assimilation techniques. Scientists are skilled at asking the right questions and they are asked to define targets that can be applied to real-world problems. AI specialists have a unique ability to `open the box' of complicated algorithms and unlock the potential of vast amounts of data. \\

Despite these challenges, scientific communities have not been deterred from exploring the interactions between ML and complex flows. On the contrary, the potential impact is attracting increasing attention, resulting in a convergence of challenges and new approaches that we believe are likely to continue transforming both fluid mechanics and machine learning research.

\acknowledgments
MB acknowledges Prof. Luca Biferale for useful discussion and financial support from the European Research Council (ERC) under the European Union’s Horizon 2020 research and innovation programme (Grant Agreement No. 882340).

%%%%%%%%%%%%%%%%%%%%%%%%%%%%%%%%%%%%%%%%%%%%%%%%%%%%%%%%
%Bibliography 
%------------

\bibliographystyle{eplbib.bst}
\bibliography{biblio}

\end{document}